\documentclass[sigconf]{acmart}

\acmYear{2026}
\copyrightyear{2026}
\acmConference[FSE Companion '26]{34th ACM Joint European Software Engineering Conference and Symposium on the Foundations of Software Engineering}{July 5--9, 2026}{Montreal, QC, Canada}
\acmBooktitle{34th ACM Joint European Software Engineering Conference and Symposium on the Foundations of Software Engineering, July 5--9, 2026, Montreal, QC, Canada}
\acmPrice{}
\acmDOI{}
\acmISBN{}

\usepackage{algorithmic}
\usepackage{graphicx}
\usepackage{booktabs}
\usepackage{float}
\usepackage{caption}
\usepackage{xcolor}
\usepackage{url}



\begin{document}

\title{A Vision for Context-Aware CI Adoption Decisions}

\author{Osamah H. Alaini}
\orcid{0009-0009-1018-1822}
\affiliation{
  \institution{Trent University}
  \department{Department of Computer Science}
  \city{Peterborough}
  \state{Ontario}
  \country{Canada}
}
\email{osamahalaini@trentu.ca}

\author{Taher A. Ghaleb}
\orcid{0000-0001-9336-7298}
\affiliation{
  \institution{Trent University}
  \department{Department of Computer Science}
  \city{Peterborough}
  \state{Ontario}
  \country{Canada}
}
\email{taherghaleb@trentu.ca}

\begin{abstract}
Continuous Integration (CI) is widely adopted in modern software development, yet adoption decisions are often made without systematic consideration of project context. Platforms such as GitHub Actions lower the barrier to CI adoption but provide limited support for grounding adoption decisions in project characteristics, leading to redundant services, unmaintained workflows, and costly migrations. Existing research and tooling primarily focus on improving CI after adoption, offering little guidance for assessing suitability before adoption. As a result, CI is frequently treated as universally beneficial rather than context-dependent.
This paper envisions a shift from default CI adoption to deliberate, context-aware decision-making. We propose an AI-enabled framework that assesses whether projects are likely to benefit from CI, recommends suitable CI services based on project characteristics, and provides configuration guidance tailored to project needs. We outline a research agenda combining developer studies, large-scale repository mining, and recommendation system design to enable informed CI adoption decisions and prevent inefficiencies before they occur.
\end{abstract}

\maketitle

\vspace{-2pt}
\section{Introduction}
\vspace{-1pt}
Continuous Integration (CI) adoption in open-source projects has surpassed 50\%~\cite{golzadeh2022rise}, yet 23\% of these adoptions are later abandoned or become obsolete~\cite{chopra2025multiCI}, and configuration issues account for 18.22\% of GitHub Actions questions on Stack Overflow~\cite{zhang2024developers}. This paradox reveals a core problem: CI adoption is often reflexive rather than deliberate. Platforms like GitHub Actions offer one-click integration without assessing project suitability (Figure~\ref{fig:vision}, top), resulting in redundant services, unmaintained workflows~\cite{chopra2025multiCI,valenzuela2024hidden}, and costly trial-and-error configuration~\cite{delicheh2026automation}.
Prior research reports that developers struggle with CI decisions. Beyond the multi-CI adoption complexity and 23\% abandonment rate documented in Java projects~\cite{chopra2025multiCI}, 52\% of developers desire easier configuration~\cite{hilton2017tradeoffs}, and complex setups remain persistent barriers~\cite{pinto2018practices,widder2019painpoints,delicheh2026automation,abrokwah2025complexity}. Despite the emergence of large language model (LLM)-based tools that enable generating CI configurations~\cite{ghaleb2025llm4ci,hossain2025cigrate}, developers still face challenges in deciding whether and how to adopt CI. Researchers have studied adoption barriers~\cite{hilton2016usage}, service migration patterns~\cite{rostami2023usage}, and CI optimization~\cite{gallaba2018use}. However, existing work focuses on post-adoption optimization rather than pre-adoption assessment, treating CI suitability as a given instead of a question worth asking.

We argue that \textit{project context} (the characteristics determining CI needs) should guide three decisions: (1) whether to adopt CI, (2) which service to use, and (3) how to configure it. Relevant context includes code complexity, testing maturity, team size, commit frequency, and resource constraints. Considering these factors turns CI adoption from default choices into deliberate, evidence-based decisions. This paper presents a vision for context-aware CI adoption that replaces reflexive defaults with intelligent assessment. We propose an AI-enabled framework that (1) evaluates whether projects need CI, (2) recommends services aligned with project characteristics, and (3) offers configuration guidance tailored to context. Drawing on recommendation systems for software engineering~\cite{robillard2009recommendation} and explainable AI~\cite{tantithamthavorn2021actionable}, we outline a research agenda to (i) survey developers about decision factors, (ii) mine repositories to link contextual factors with adoption outcomes, and (iii) build an empirically grounded recommendation framework.

\vspace{-3pt}
\section{Why Assess CI Adoption Suitability?}
\label{sec:why}

\vspace{-2pt}
\subsection{\large Motivating Scenarios}
\vspace{-1pt}
Consider the following three representative scenarios that illustrate the need for context-aware CI assessment:

\vspace{2pt}
\noindent\textit{\textbf{Scenario 1: Over-adoption.}} A personal portfolio website with monthly updates adopts GitHub Actions with build, test, and deployment workflows. The repository contains no tests, receives 2-3 commits per month from a single developer, and has zero production dependencies. CI provides minimal value but incurs configuration maintenance costs and failed workflow notifications.

\vspace{2pt}
\noindent\textit{\textbf{Scenario 2: Under-adoption.}} A financial services library used by 50+ internal apps lacks CI. The repository receives daily commits from a 15-person team and has critical authentication logic with 80\% test coverage. Developers run tests manually, leading to integration failures in downstream systems. CI would be highly valuable but is not adopted due to perceived configuration complexity.

\vspace{2pt}
\noindent\textit{\textbf{Scenario 3: Service mismatch.}} A machine learning research project uses Travis CI configured for standard Python unit tests. The project requires GPU-accelerated model training, large dataset downloads, and matrix builds across multiple framework versions. GitHub Actions with custom runners would better suit these needs, but the team lacks guidance for migration.

\vspace{4pt}
\noindent Our vision addresses these scenarios through proactive assessment (preventing \textit{Scenario 1}), recommendation (assessing \textit{Scenario 2}), and service matching (resolving \textit{Scenario 3}).

\vspace{-2pt}
\subsection{\large Reflection on Research}
\vspace{-2pt}
\textbf{\textit{Research analyzes CI retrospectively, not proactively.}}
Prior research reports CI benefits and challenges but focuses on \textit{how} to implement CI, not \textit{whether} to adopt it. Outcomes are assessed only after resources are committed, leaving a gap in proactive guidance. Recommendation systems for software engineering (RSSEs) help developers make complex decisions~\cite{robillard2009recommendation}, yet 91\% are reactive and focus on code-level tasks such as reuse, debugging, or API selection~\cite{gasparic2016recommendation}. None assesses infrastructure choices, including CI suitability.
Even recent LLM-based tools for generating CI configurations~\cite{ghaleb2025llm4ci,hossain2025cigrate,ghaleb2026agentscicd} focus on setup rather than on deciding whether CI should be adopted. 
Prior work shows widespread uninformed CI adoption. In 18,924 Java projects, 23\% of CI adoptions are abandoned or become obsolete, and 18\% use multiple services, often due to unplanned migrations~\cite{chopra2025multiCI}. Configuration complexity is central~\cite{widder2019painpoints,pinto2018practices}: in 2,557 Android projects, 81\% of CI/CD effort targets setup and only 9\% deployment~\cite{ghaleb2025cicd}, contributing to abandonment. Workflow files require ongoing maintenance~\cite{valenzuela2024hidden}, and 71\% of developers lack tools to monitor CI practices~\cite{santos2025monitoring}, hindering assessment of alignment with project needs.
These findings are valuable but retrospective. Developers still lack frameworks to assess whether CI justifies its overhead for their team size, commit frequency, test coverage, and code complexity, or to choose a suitable service. Existing work focuses on fixing adoption mistakes rather than preventing them. We argue that understanding \textit{when} and \textit{why} CI suits a project is as important as knowing \textit{how} to implement it.

\vspace{-2pt}
\subsection{\large Reflection on Practice}
\vspace{-2pt}
\textbf{\textit{Practice shows adoption without assessment.}}
On platforms like GitHub, CI adoption is easy but often uninformed (Figure~\ref{fig:vision}, top). GitHub Actions lets any repository enable CI with a single click, without considering project-specific needs. This broad access to automation creates systematic waste.
Prior research shows projects abandoning CI after adoption~\cite{chopra2025multiCI}, using multiple CI services~\cite{chopra2025multiCI}, and struggling with configuration~\cite{hilton2017tradeoffs,pinto2018practices,widder2019painpoints}, which harms both adoption and build performance~\cite{ghaleb2019duration}. Adoption is often driven by social influence, platform defaults, or bandwagon effects rather than careful project evaluation.
Teams with legacy systems face high migration costs to make codebases CI-compatible, yet GitHub Actions promotes uniform adoption. Misconfigurations are common: Khatami et al.~\cite{khatami2024catching} found 36.5\% of changes adjust run steps, and Delicheh et al.~\cite{delicheh2026automation} report practitioners relying on trial-and-error, with one needing ``\textit{30 builds or releases to tune it to work as expected}''.
Incorrect adoption accumulates technical debt: abandoned configurations, redundant services, and costly migration rework. Across 28,770 repositories, build (80\%) and test (80.2\%) stages are common, but static application security testing (12.1\%) and cloud deployment (4.2\%) are rare. CI usage also differs: Travis CI dominates in building (99\%), whereas GitHub Actions in linting (45\%), yet developers still lack frameworks to match services to project context~\cite{chomatek2025decoding}.

\smallskip
\noindent
\textbf{Summary.}  
CI adoption via platforms like GitHub Actions has become easy but often not deliberate. Developers can enable CI with little effort, yet developers rarely assess whether CI suits their project. This leads to abandoned configurations, recurring setup issues, and repeated trial-and-error. Decisions are often driven by defaults, trends, or perceived best practices rather than careful evaluation, resulting in misconfigured pipelines, underused features, and poorly matched services that accumulate into technical debt and unnecessary complexity.

\vspace{-3pt}
\section{A Vision of Context-Aware CI Adoption}
\label{sec:vision}
\begin{figure*}[t]
    \centering
    \vspace{-1pt}
    \includegraphics[width=.71\linewidth]{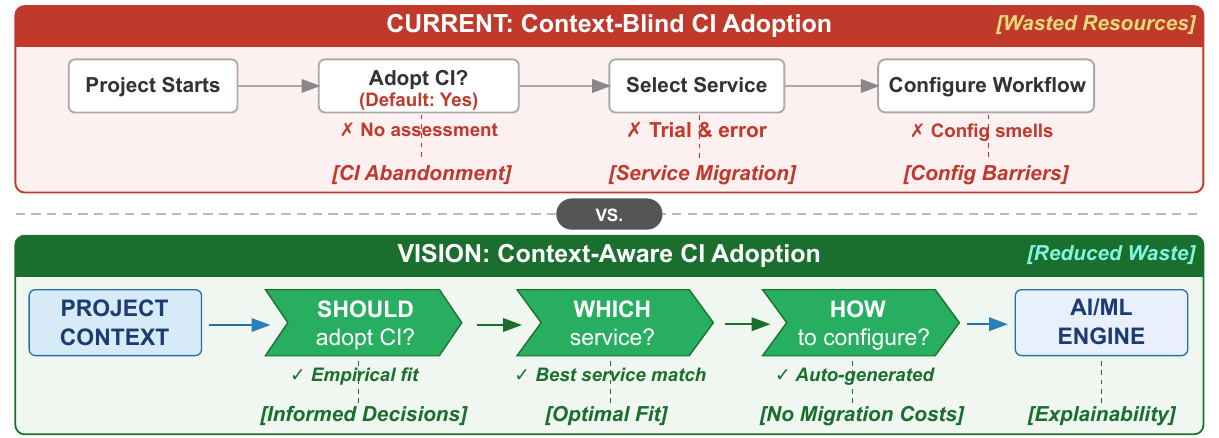}
    \vspace{-4pt}
    \caption{Current context-blind CI adoption (top) versus our envisioned 
    context-aware framework (bottom).}
    \label{fig:vision}
    \vspace{-7pt}
\end{figure*}

\vspace{-2pt}
Figure~\ref{fig:vision} contrasts current context-blind CI adoption (top) with our vision of context-aware decisions (bottom). Instead of immediately enabling GitHub Actions, a team planning CI for an open-source project consults an \textbf{\textit{AI-enabled recommendation system}} that analyzes commit frequency, test coverage, team size, code complexity, and deployment needs. It then determines: (1) whether CI benefits the project, (2) which CI service best suits its needs, and (3) how to configure it based on empirical best practices. This shifts CI adoption from defaults to evidence-based guidance to answer: \textit{"Does my project need CI, and if so, which approach suits my context?"}

\vspace{4pt}
\noindent Our vision rests on three AI-enabled components:

\vspace{-5pt}
\subsection{\large Project Context Model}
\vspace{-1pt}
The foundation captures contextual factors that determine CI suitability and service fit, including development activity (commit frequency, contributor count, collaboration intensity), codebase characteristics (languages, dependencies, test coverage), project maturity (age, maintenance status, community size), quality assurance (testing infrastructure, code review), and resource constraints (compute, budget, expertise). Large-scale studies show distinct adoption patterns: Chomtek et al.~\cite{chomatek2025decoding} report that 80\% of projects use basic build/test automation, but only 4.2\% deploy to the cloud, indicating CI should scale with project needs rather than follow a universal template. The context model thus bases adoption decisions on empirical evidence rather than assumptions or social influence.

\vspace{-2pt}
\subsection{\large AI-Enabled Suitability Assessment}
\vspace{-1pt}

Machine learning (ML) models can predict whether projects benefit from CI adoption and 
which service aligns with their characteristics. Unlike existing approaches that assume universal CI benefits or analyze failures retrospectively, our framework operates \textit{proactively} before developers commit resources.
Prior research shows 82\% of software practitioners consider it as important as prediction accuracy~\cite{jiarpakdee2021practitioners}. Developers will not trust opaque assessments. While rule-based heuristics such as \textit{adopt CI if commit frequency~$>$5/week} are simple, adoption outcomes vary substantially across project types even at similar commit frequencies~\cite{chopra2025multiCI,hilton2016usage}, and fixed thresholds fail to generalize across diverse domains~\cite{rostami2023usage,chomatek2025decoding}.
Static rules also become stale as CI ecosystems evolve. Our Stage~1 survey will empirically map the heterogeneous factors developers use when making CI decisions, validating that these interactions resist reduction to static rules and informing the feature space for ML models.

The assessment addresses three questions: (1) Is CI necessary given the project context? (2) Which CI service best suits the project's needs? (3) What practices are suitable versus unnecessary complexity? This prevents costly trial-and-error. Delicheh et al.~\cite{delicheh2026automation} documented practitioners requiring ``\textit{30 builds or releases to tune it to work as expected}'', and approximately 23\% of CI adoptions are eventually abandoned or obsolete~\cite{chopra2025multiCI}, leaving unmaintained configuration debt. Stage~2 will mine large-scale repositories to correlate project characteristics with adoption outcomes.

\vspace{-2pt}
\subsection{\large AI-Powered CI Recommendations and Configurations}
\vspace{-1pt}

Generative AI, including LLMs, turns suitability assessments into actionable guidance by providing personalized configurations and clear explanations that surpass static recommendations~\cite{tantithamthavorn2021actionable}. Developers receive ranked service recommendations with context-aware justifications, effort and complexity estimates, auto-generated workflows based on empirically validated practices, and advice on which practices to adopt now versus later. For projects that are poor CI candidates, AI explanations clarify why adoption may waste resources and when CI could become beneficial, helping avoid the 23\% abandonment rate~\cite{chopra2025multiCI}.
Our vision is to give developers intelligent assessments before committing resources, supported by extensive empirical data, sophisticated modeling, validation across diverse projects, and continuous updates as CI ecosystems evolve. Intermediate steps toward this vision (detailed in Section~4) already offer substantial benefits: preventing uninformed adoption, reducing configuration complexity through evidence-based guidance, and minimizing wasted resources. CI adoption becomes a deliberate, AI-informed decision grounded in project context and aligned with the rigor of other software engineering practices.

\vspace{-3pt}
\section{Research Agenda}
\label{sec:agenda}
\vspace{-2pt}

Realizing context-aware CI adoption requires addressing three fundamental research challenges. We organize our agenda into three stages that build toward the AI-enabled framework described in Section~\ref{sec:vision}, where each stage produces actionable insights while laying the foundation for subsequent work.

\vspace{-3pt}
\subsection{\large Stage 1: Understanding CI Adoption Drivers and Decision Factors}
\vspace{-2pt}
Before designing context models or assessment frameworks, we will investigate how developers make CI adoption decisions, what influences these choices, and whether they would trust AI-based guidance. We will combine large-scale surveys of GitHub developers with targeted interviews to explore: (1) what drives CI adoption decisions, (2) what factors determine CI service selection, (3) how developers assess project–CI fit, and (4) what makes AI recommendations trustworthy in practice.

These findings will guide the design of the AI-powered system. If developers rely on comparing similar projects, the system can formalize that process using empirical evidence. If cost is an impact factor, recommendations will highlight pricing tradeoffs. Insights about trust will further inform how explanations are presented to ensure informed adoption.  

The study will also ground the return on investment (ROI) model introduced later, capturing the tradeoff between CI benefits and adoption or maintenance costs. Survey responses on perceived effort, configuration difficulty, and maintenance burden will serve as proxies for cost terms, while benefit and usage frequency will be derived empirically in subsequent stages through repository mining and CI-related quality metrics.

\vspace{-3pt}
\subsection{\large Stage 2: Extracting and Correlating Contextual Factors}
\vspace{-2pt}
Intuitions about which project characteristics justify CI adoption require empirical validation. We will systematically analyze repositories to identify contextual factors that correlate with CI adoption success, abandonment, or maintenance burden.
Mining large-scale repository data will allow us to: (1) extract project characteristics such as development activity, codebase metrics, team structure, and testing practices, (2) track CI adoption patterns over time, including adoption, abandonment, migration, and configuration evolution, (3) identify correlations between context and outcomes, such as predictors of successful adoption or indicators of maintenance burden, and (4) discover service selection patterns to see which project types gravitate toward specific CI services. Processing repositories at scale and applying statistical methods will reveal significant correlations to inform AI training.
This analysis provides immediate value, highlighting actionable insights such as relationships between commit frequency and abandonment rates. More importantly, it generates the empirical ground truth needed for AI-enabled recommendation systems. Predictive models cannot be built without linking project characteristics to adoption outcomes, and understanding these correlations also tests assumptions such as ``\textit{All projects benefit from CI}'', establishing which contexts well justify investment.

\vspace{-3pt}
\subsection{\large Stage 3: Building AI-Enabled Recommendation Framework}  
\vspace{-2pt}
Our goal is to create intelligent systems that assess CI suitability, recommend suitable services, and generate configurations. Building these systems requires careful modeling, extensive training data, validation across diverse projects, and interfaces developers trust. Challenges in software engineering contexts, including limited baseline systems, data constraints, and complex evaluation, demand rigorous empirical design and validation.

The AI-enabled framework will implement three capabilities:
\vspace{-4pt}
\begin{enumerate}
    \item \textit{Context extraction and modeling}: We will automatically extract project characteristics from repository data in ML-ready formats. This includes feature engineering to capture relevant signals, temporal modeling to track project evolution, and handling evolving contexts as projects mature.

    \item \textit{ML-based suitability classification}: We will train models to predict whether projects benefit from CI, which practices are suitable, and abandonment risk based on project characteristics. We will balance precision and recall, starting with interpretable models (e.g., logistic regression) and, if warranted, moving to neural networks. Suitability scores will compare CI's benefits and costs. Benefits are estimated as time saved by automated tests across runs; costs as effort to set up and maintain CI. We will derive these from survey responses, abandonment rates, configuration overhead, and build and quality metrics. Projects with higher estimated benefit than cost will be recommended to adopt CI; if values are close, we will issue an \emph{uncertain} recommendation with explanations of the main cost drivers.
    
    \item \textit{AI-enabled CI service recommendations and configurations}: We will develop ranking algorithms to match project requirements with CI service capabilities, considering language support, features, and cost. Generative AI will create initial workflow configurations based on patterns from similar successful projects, reducing trial and error. Natural language explanations will justify recommendations, e.g., ``\textit{CI recommended: your commit frequency (15/week) and team size (8 contributors) match patterns from successful CI adopters with high retention rates}''.

\end{enumerate}

\vspace{-4pt}
\subsection{\large Evaluation Plan}
\vspace{-3pt}
\smallskip\noindent\textbf{Ground truth.}
We will evaluate Stage~3 models on a held-out subset of the Stage~2 corpus, targeting over 5,000 repositories in Java~\cite{chopra2025multiCI}, JavaScript~\cite{golzadeh2022rise}, and Android~\cite{ghaleb2025cicd,zhou2026ciadoptionmobile,parsazadeh2026instrumentation} to ensure cross-domain generalizability. Ground-truth labels will use observed adoption outcomes, including CI abandonment within 12~months and configuration churn exceeding 50\% of commits~\cite{khatami2024catching,chopra2025multiCI}.

\smallskip\noindent\textbf{Baselines and metrics.}
Baselines can include: (i) an \emph{always-adopt} rule, (ii) a single-feature threshold classifier (e.g., \textit{commit frequency $>$10/week}), and (iii) a popularity-based recommender defaulting to the most prevalent CI per language. Success criteria are F1\,$\geq$\,0.75 and AUC-ROC\,$\geq$\,0.80 for suitability classification, NDCG@3\,$\geq$\,0.70 for service ranking, and perceived explanation usefulness\,$\geq$\,80\% validated through a follow-up practitioner survey~\cite{jiarpakdee2021practitioners}.

\smallskip\noindent\textbf{Explainability.}
Developers may not trust assessments of CI suitability if reasoning is opaque. AI-generated explanations can clearly show why a suggestion suits the project context, what risks it entails, and under what conditions it might change. This transparency addresses the common distrust of black-box or ``\textit{AI magic}'' systems. We plan to assess whether AI-generated explanations improve developer trust and understanding, as measured by perceived usefulness in a follow-up survey or through pull-requests.

\vspace{-2pt}
\subsection{\large Potential Challenges}
We anticipate three main challenges in building context-aware CI recommendation systems:

\smallskip\noindent\textbf{Data availability.}
Training accurate models requires large-scale data on CI adoption outcomes, including reasons for abandonment and configuration evolution, much of which is missing from public repositories. Early-stage projects with limited historical activity present a cold-start challenge. We will address this using lightweight onboarding proxies, such as declared language, dependency manifest complexity, and stated team size, which enable initial recommendations before longitudinal data accumulates.

\smallskip\noindent\textbf{Bias and generalizability.}
Training on open-source repositories risks popularity bias, as GitHub Actions and Travis CI dominate the corpus~\cite{chopra2025multiCI,golzadeh2022rise}. We will mitigate this through stratified sampling across CI services, project domains, and team sizes to ensure recommendations reflect optimal fit rather than prevalence. Generalizability across different domains, including web applications, embedded systems, and data science, remains uncertain and will require careful validation.

\smallskip\noindent\textbf{Evolving ecosystems and explainability.}
Changes in CI landscape can render models trained on historical data obsolete without continuous updates. In addition, explainability is essential for developer trust. Recommendations must clearly show why a suggestion suits the project context, the risks involved, and conditions under which it might change, allowing developers to maintain autonomy while using AI guidance.

\vspace{-1pt}
\section{Conclusion}
\label{sec:conclusion}
\vspace{-1pt}
CI is often adopted without considering project context. Many projects enable GitHub Actions simply because it is free and integrated, not because it suits their needs. This can create waste: abandoned workflows, redundant services, and repeated configuration changes that consume development time. Evidence shows 23\% of adoptions are abandoned or obsolete, 36.5\% of commits modify CI configurations, and some workflows require up to 30 builds to tune.
Our vision challenges the assumption that CI is universally beneficial. Not all projects need CI, and those that do may require different approaches. Recommending CI without considering project characteristics can create inefficiency. Our research agenda moves from post hoc analysis to proactive guidance by studying developer motivations, correlating project context with CI outcomes, and building an AI-enabled recommendation framework.  
Achieving effective CI adoption requires empirical, context-aware decisions. This vision provides a pathway to align CI adoption with project needs and prevent waste before it occurs.

\vspace{-1pt}
\begin{acks}
\vspace{-1pt}
This work is funded by the Natural Sciences and Engineering Research Council of Canada (NSERC): RGPIN-2025-05897.
\end{acks}

\clearpage
\balance
\bibliographystyle{ACM-Reference-Format}
\bibliography{main}

\end{document}